\documentclass[12pt]{iopart}
\expandafter\let\csname equation*\endcsname\relax
\expandafter\let\csname endequation*\endcsname\relax
\usepackage{graphicx,color,aas_macros,amsmath}
\begin{document}
\title{Thermodynamically-consistent semi-classical $\ell$-changing rates}
\author{R.~J.~R.~Williams$^1$, F.~Guzm\'an$^2$, N.~R.~Badnell$^3$,
  P.~A.~M.~van~Hoof$^4$, M.~Chatzikos$^2$, G.~J.~Ferland$^3$
}
\address{$^1$ AWE plc, Aldermaston, Reading RG7 4PR, UK}
\address{$^2$ Department of Physics and Astronomy, University of Kentucky, Lexington, KY 40506, USA}
\address{$^3$ Department of Physics, University of Strathclyde, Glasgow G4 0NG, UK}
\address{$^4$ Royal Observatory of Belgium, Ringlaan 3, 1180 Brussels, Belgium}

\date{\today}
\begin{abstract}
  We compare the results of the semi-classical (SC) and
  quantum-mechanical (QM) formalisms for angular-momentum changing
  transitions in Rydberg atom collisions given by Vrinceanu \&
  Flannery, J.\@ Phys.\@ B {\bf 34}, L1 (2001), and Vrinceanu, Onofrio
  \& Sadeghpour, ApJ {\bf 747}, 56 (2012), with those of the SC
  formalism using a modified Monte Carlo realization.  We find that
  this revised SC formalism agrees well with the QM results.  This
  provides further evidence that the rates derived from the QM
  treatment are appropriate to be used when modelling recombination
  through Rydberg cascades, an important process in understanding the
  state of material in the early universe.  The rates for
  $\Delta\ell=\pm1$ derived from the QM formalism diverge when
  integrated to sufficiently large impact parameter, $b$.  Further to
  the empirical limits to the $b$ integration suggested by Pengelly \&
  Seaton, MNRAS {\bf 127}, 165 (1964), we suggest that the fundamental
  issue causing this divergence in the theory is that it does not
  fully cater for the finite time taken for such distant collisions to
  complete.
\end{abstract}

\pacs{32.80.Ee, 34.10.+x, 34.50.Fa}
\submitto{\jpb}
\maketitle

\section{Introduction}
\label{intro}

There has been significant recent interest in the rates for
$\ell$-changing collisions in Rydberg atoms.  While this may seem a
somewhat obscure corner of atomic physics, the time to pass through
the ladder of high-angular momentum levels in highly excited atoms
proves to be a bottleneck in the process of atomic (re-)combination in
the early universe.  The details of the $\ell$-changing rates
therefore have a major impact on our understanding of this important
stage in cosmic development \cite{2010MNRAS.407..599C}.

The atomic physics of these rates can be calculated by a variety of
approximations, as described in detail by \cite{Vrinceanu2001b} and
\cite{VOS2012} (hereafter VF01 and VOS12, respectively).  In this
paper, we will concentrate on the differences between the most
detailed of these formulations, those based on quantum-mechanical
perturbation theory (QM) and semi-classical trajectory theory (SC).
As discussed elsewhere
(\cite{2016MNRAS.459.3498G,2017MNRAS.464..312G}, and references
therein), these theories result in predictions for integrated rates
which differ by up to an order of magnitude, sufficient to have a
major impact on the interpretation of observations.  This difference
is primarily the result of the different range in impact parameter
over which $\vert\Delta\ell\vert = 1$ transitions are active in the
different theories: the QM theory has a logarithmic divergence in
transition rate, which requires a limit to be placed on the largest
impact parameter for which collisions are active in causing
transitions, while in the VF01/VOS12 SC approach, this limit arises as
a result of the finite domain in impact parameter over which a single
collision can cause a complete $\vert\Delta\ell\vert = 1$ transition.

In this paper, we will present the results of calculations using the
semi-classical theory of VF01, but using an alternative Monte Carlo
formalism.  We find that with this alternative approach, the classical
results agree well with those of the quantum theory.  The differences
which there are, are of the kind that would usually be expected when
relating a quantum theory to its classical approximation, and are
consistent with the correspondence principle.  When integrated over
impact parameter, $b$, and a thermal spectrum of colliders, it is
clear that the differences in rates computed using this modified SC
approach and the original QM formalism will be minimal.

In Section~\ref{s:calc} we compare the quantum mechanical transition
probabilities with semi-classical probabilities as calculated by VF01
and VOS12, and with our revised approach.  In Section~\ref{s:use}, we
briefly discuss how the discretely-sampled VF01 SC approach can be
made somewhat more accurate, and consistent with the discrete detailed
balance relations.  Finally, in Section~\ref{s:concl}, we summarize
our results, and discuss the processes which prevent the overall
dipole transition rate from diverging in a plasma of finite density.

\section{Calculations}
\label{s:calc}
VF01 and VOS12 provide detailed formulae for the rates of transitions
$\ell\to\ell'$ for precise values of $\ell$, $\ell'$.  They derive
their semi-classical formulae using approximations which correspond to
the continuum limit $n,\ell,\ell' \to \infty$, with $\ell/n$ and
$\ell'/n$ finite, and then apply them in the case of finite quantum
numbers.  They apply what they term as a microcanonical ensemble,
sampling the fundamentally continuous classical-limit expressions at
discrete values of incoming and outgoing angular momentum appropriate
for the angular momentum quantum number.  Using this procedure
provides values for the overall collision rate which are finite for
all $\Delta\ell$, as noted above.

The process of returning from the continuum to the discrete limit is
not, however, unique.  In the present paper, we use a method similar
to that discussed by \cite{1984JPhB...17.3923B}.  To ensure
thermodynamic consistency for the derived total rates, it is better to
follow a finite-volume formalism (see, e.g.,
\cite{leveque2002finite}), where each of the discrete quantum numbers
is taken to correspond to a finite range of continuum values.  The
simplest assumption which will ensure results are consistent with the
thermodynamical equilibrium is to assume that the probability density
of states in the continuum band corresponding to each of the aggregate
states is internally in thermodynamic equilibrium.  As there is no
energy difference between states in the case considered by Vrinceanu
et al., this corresponds to assuming a uniform population.  This
procedure provides results consistent with the thermodynamic
requirements of unitarity and detailed balance and with the quantum
and classical limits, as well as with usual practice in Monte Carlo
simulation of off-lattice systems \cite{landau2014guide}.  However, it
leaves us looking to statistical physics, rather than numerical
convention, to resolve the divergence in dipole transition rates.

For the standard quantum mechanical association of the radial quantum
number $n$, the total angular momentum quantum number $\ell$ satisfies
$0\le \ell \le n-1$. If we assume that each value of $\ell$ maps to a
shell with total angular momentum between $\ell \hbar$ and $(\ell+1)
\hbar$, with a classical density of states $\propto\ell$ (which may be
visualized as a two-dimensional polar coordinate system), then the
area of this shell is $\propto 2\ell+1$.  This is consistent with the
number of $z$-angular momentum eigenstates $\vert m \vert \le \ell$
corresponding to each total angular momentum eigenstate.  It results
in a mean-squared angular momentum in the shell of
\begin{equation}
\left\langle L^2\right\rangle = \left[\ell(\ell+1)+{1\over 2}\right]\hbar^2,
\end{equation}
which is a constant $\hbar^2/2$ greater than the value which enters in
quantum mechanical calculations, $\left\langle L^2\right\rangle =
\ell(\ell+1)\hbar^2$.  For comparison, assuming that the total angular
momentum corresponding to a quantum number $\ell$ is exactly
$\ell\hbar$ underestimates $\left\langle L^2\right\rangle$ by
$\ell\hbar^2$, which is a significantly larger error for large $\ell$.
[Taking the angular momentum for the discrete state to be
  $\left(\ell+{1\over2}\right)\hbar$ is more accurate, with the
  classical and quantum density of states being equivalent, and the
  mean-square angular momentum $\left\langle L^2\right\rangle$
  over-estimated by $\hbar^2/4$.]

Transition probabilities in the finite volume regime can be derived
from the semi-classical results of VF01/VOS12 by interpreting them as
probability density functions in the continuum limit, so the
transition probability from the state $(n,\ell)$ to $(n,\ell')$
becomes
\begin{align}
  \begin{split}
  &\left\langle P^{\rm SC}\right\rangle_{n\ell\ell'} = \\
  &\qquad\int_{\ell/n}^{(\ell+1)/n} {\rm d} \lambda
  \int_{\ell'/n}^{(\ell'+1)/n} {\rm d} \lambda' 
  \, P^{\rm SC}(\lambda,\lambda',\chi)g(\lambda)g(\lambda')
  \\&\qquad\qquad\left/ \int_{\ell/n}^{(\ell+1)/n} {\rm d} \lambda \, g(\lambda)\right.,
  \end{split}
  \label{e:phase}
\end{align}
where $g(\lambda) = 2\lambda$ is the classical density of states,
normalized so that $\int_0^1 g(\lambda){\rm d}\lambda = 1$, and
\begin{align}
  \begin{split}
  P^{\rm SC}(\lambda,\lambda',\chi) &= \\
  &{2\lambda'/n\over \pi\hbar \sin\chi}\left\{ 
\begin{array}{ll}
0, & \vert \sin\chi\vert < \vert\sin(\eta_1-\eta_2)\vert \\
{K(B/A)\over \sqrt{A}} 
& \vert \sin\chi\vert > \vert\sin(\eta_1+\eta_2)\vert \\
{K(A/B)\over \sqrt{B}} 
& \mbox{otherwise}, \\
\end{array}
\right.
  \end{split}
  \label{e:vos12sc}
\end{align}
is the SC transition probability given by VOS12.  In this expression,
$K$ is the complete elliptic integral,
\begin{align}
  A &= {\sin^2\chi - \sin^2(\eta_1-\eta_2)}, \\
  B &= {\sin^2(\eta_1+\eta_2)-\sin^2(\eta_1-\eta_2)}, \\
  \cos\eta_1 &= \lambda,\\
  \cos\eta_2 &= \lambda',
\end{align}
and $\chi$ is given in terms of $n$ and other implicit parameters of
the collision by
\begin{align}
  \cos\chi &=
           {1+\alpha^2\cos\left(\sqrt{1+\alpha^2}\Delta\Phi\right)\over
             1+\alpha^2},\\
  \alpha &= {3Z_1\over 2}\left(a_n v_n\over bv\right),
\end{align}
where the swept angle $\Delta\Phi$ is assumed to be $\pi$.  
  
This procedure replaces the closed-form expressions of VF01 and VOS12
with a double integral, so is not as suitable for numerical work.
However, it seems worthwhile to compare the results with those of the
discrete interpretation in order to inform possible modifications to
the VF01/VOS12 formalism which might be made to ensure compatibility
with the limit of thermodynamic equilibrium.

There are a number of desirable properties for any set of approximate
transition probabilities.  These include unitarity, i.e.\@ that the
system must reside in one of the angular momentum states at the end of
the transition
\begin{equation}
\sum_{l'} P_{n\ell\ell'} = 1,
\end{equation}
and detailed balance, i.e.\@
\begin{equation}
(2\ell+1) P_{n\ell\ell'} = (2\ell'+1)P_{n\ell'\ell},
\end{equation}
for the quantum level degeneracy $g(\ell)=2\ell+1$.  Note that the sum
for the unitarity requirement includes the probability that the
scattering leads to no transition, $P_{n\ell\ell'}$ with $\ell' =
\ell$.  It is possible to determine this rate using the same analytic
forms as for the $\ell$-changing interactions, and this rate is
included in the plots shown below.

The symmetry of the expressions for $A$ and $B$ in $\lambda$ and
$\lambda'$, together with the overall factor of $\lambda'$ in
equation~(\ref{e:vos12sc}), means that equation~(\ref{e:vos12sc})
satisfies the detailed balance relations {\em in the continuum limit},
\begin{equation}
  2\lambda P^{\rm SC}(n,\lambda,\lambda')
  =
  2\lambda'P^{\rm SC}(n,\lambda',\lambda),
\end{equation}
given the classical density of states $g(\lambda) = 2\lambda$.  As a
result of this, it is simple to verify that the phase-space average,
equation~(\ref{e:phase}), satisfies the discrete detailed balance
relation
\begin{equation}
  (2\ell+1)\left\langle P^{\rm SC}\right\rangle_{n\ell\ell'}
  =
  (2\ell'+1)\left\langle P^{\rm SC}\right\rangle_{n\ell'\ell}.
\end{equation}

Beyond these absolute requirements, we also suggest that the rates
should be subject to another statistical requirement for collisions at
small impact parameter.  For these scatterings, the output state of
the interaction is dependent on complex interference phenomena,
sensitive to many details of the atomic physics.  However, the net
effect of this complexity, when averaged over some small range of
incoming particle properties, would be expected to be asymptotically
close to the output states being in statistical equilibrium (cf., for
the classical case, \cite{1967MNRAS.136..101L}).  We therefore suggest
that, in the limit of close scatterings $b\to 0$, the rates should be
subject to an ergodicity property
\begin{equation}
\left\langle P\right\rangle_{n\ell\ell'} \simeq {2 \ell'+1\over n^2},
\end{equation}
i.e.\@ when the collider passes close enough to the core of the target
atom, the effect of the collision is to randomize the output state,
when the input state is coarse-grained over a suitable domain.  Of
course, in reality scatterings will cease to be purely $\ell$-changing
in this limit.  Even so, it is to be expected that the output state
angular momentum will become statistically independent within the
shell.  This requirement seems to be the best physical interpretation
which can be put on the statement in \cite{1964MNRAS.127..165P},
hereafter PS64, that in the core the scattering probability becomes a
rapidly-oscillating function with mean value $1\over2$.  This is what
would result from the core ergodicity principle in the case of a
two-level system, so the core ergodicity principle seems like a
reasonable generalization, agreeing with the work of PS64 at least in
spirit.  As we will see, it is also a reasonable description of what
in fact happens when the quantum and shell-averaged classical
transition probabilities are calculated in detail.

\begin{figure*}
\begin{center}
\begin{tabular}{cc}
\includegraphics[width=0.4\textwidth]{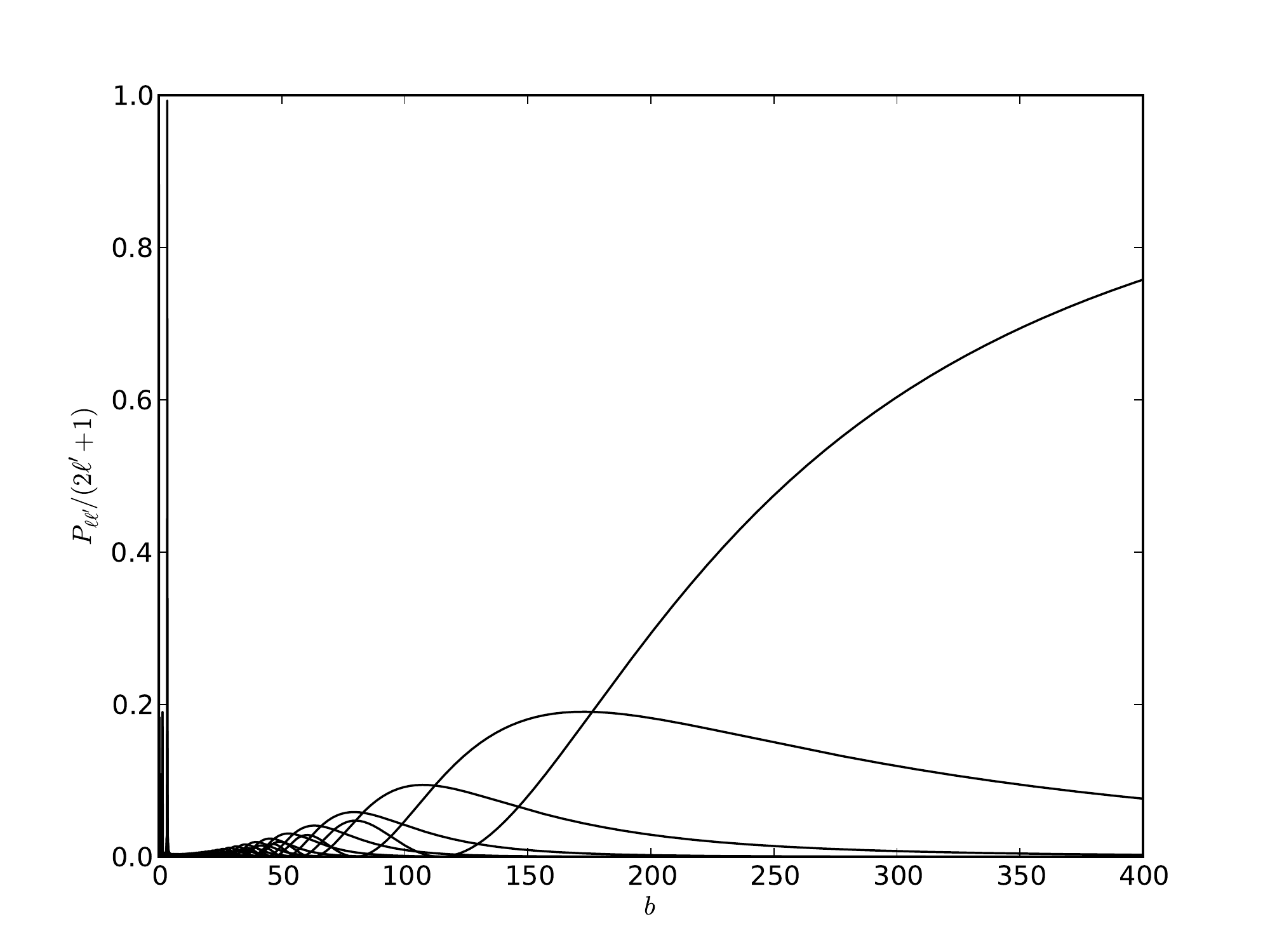} &
\includegraphics[width=0.4\textwidth]{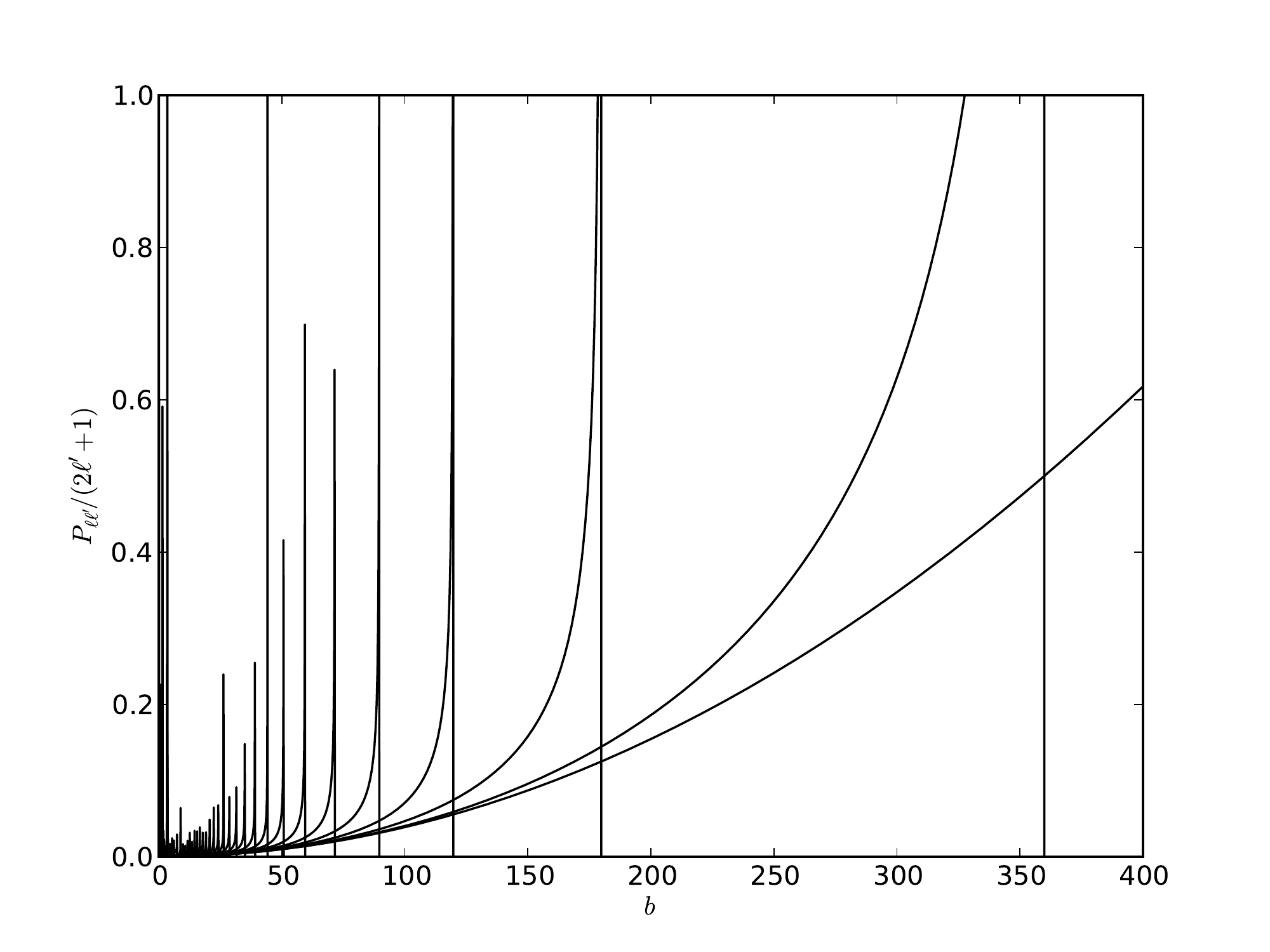} \\
\includegraphics[width=0.4\textwidth]{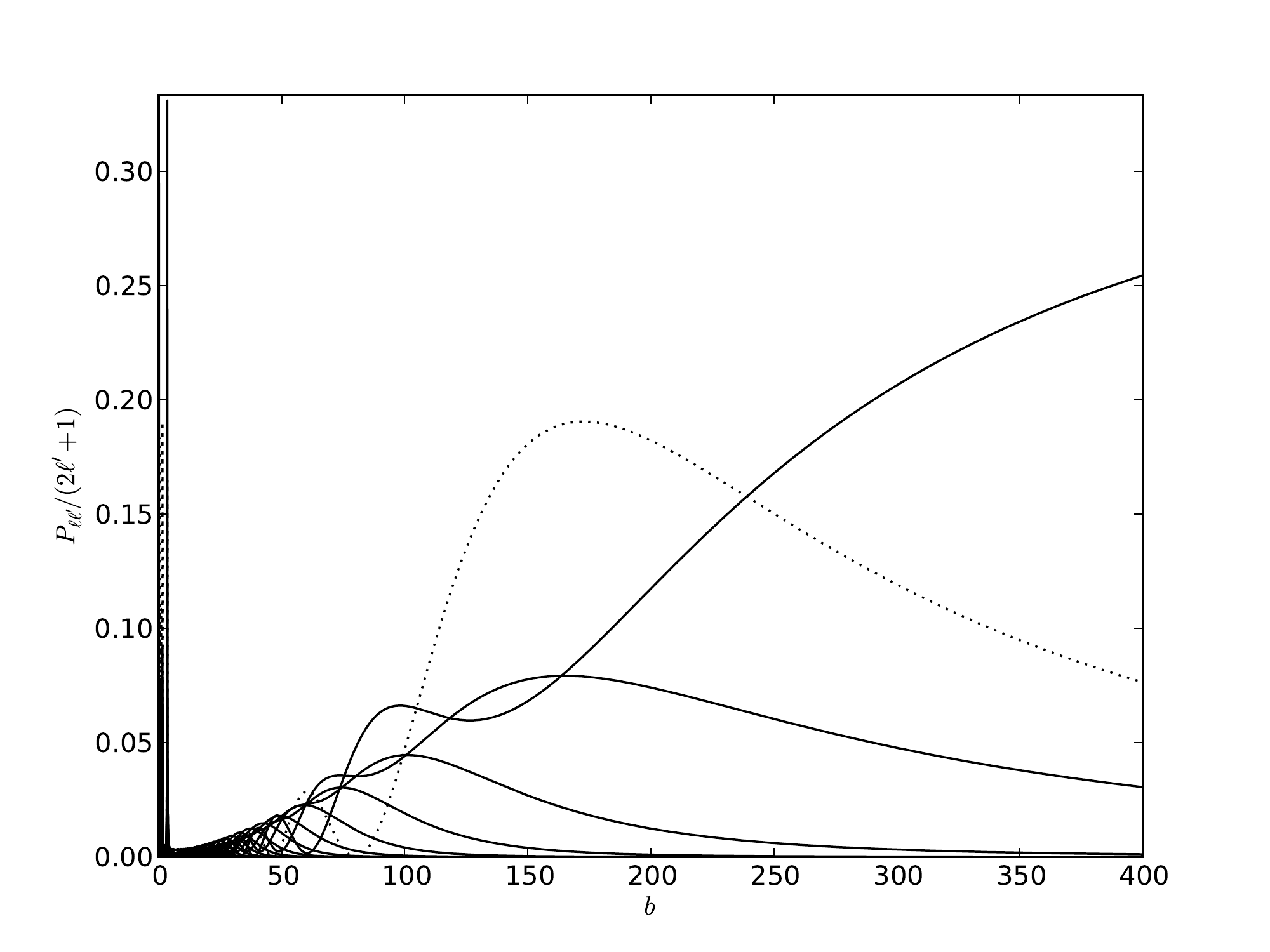} &
\includegraphics[width=0.4\textwidth]{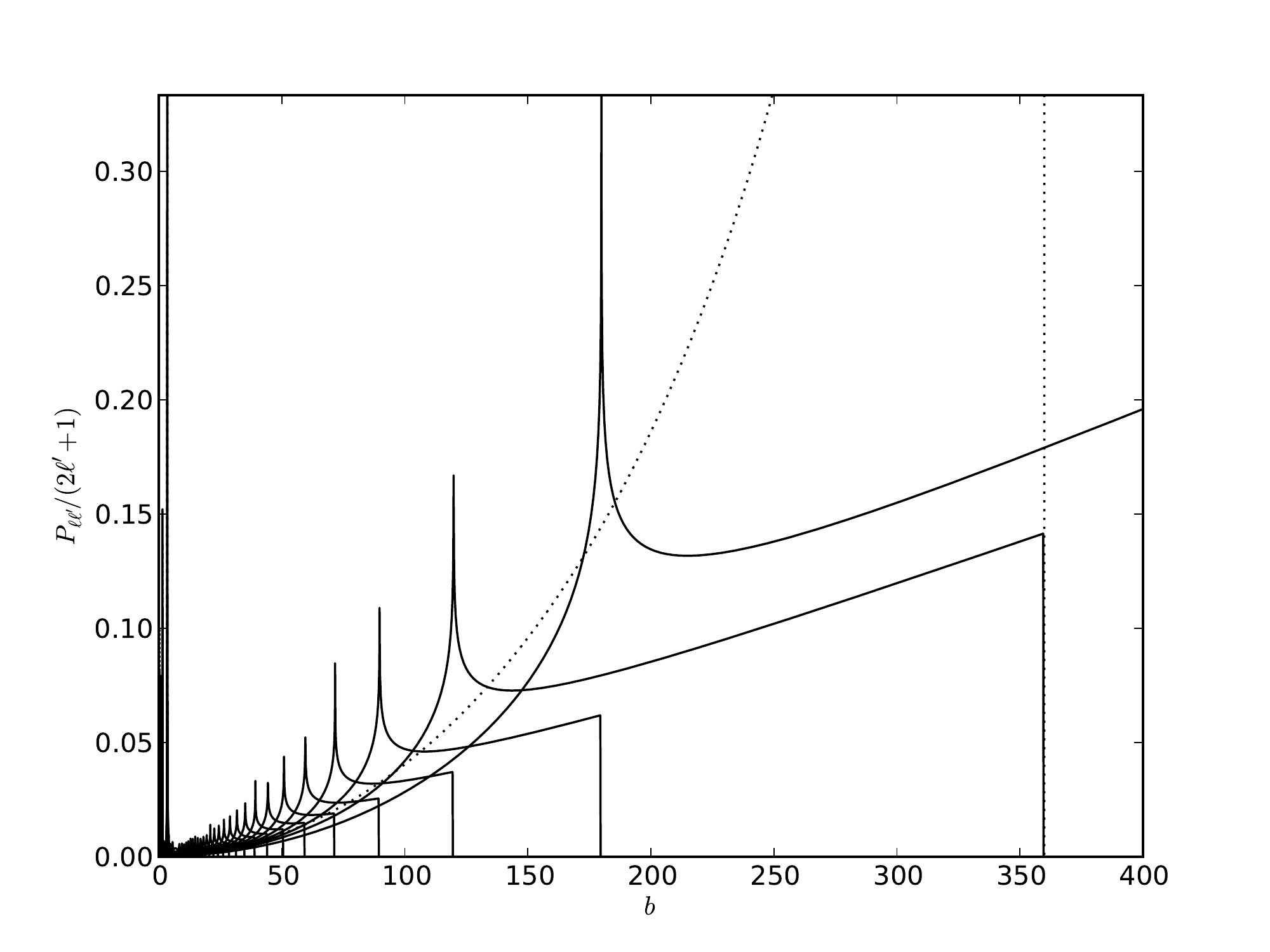} \\
\includegraphics[width=0.4\textwidth]{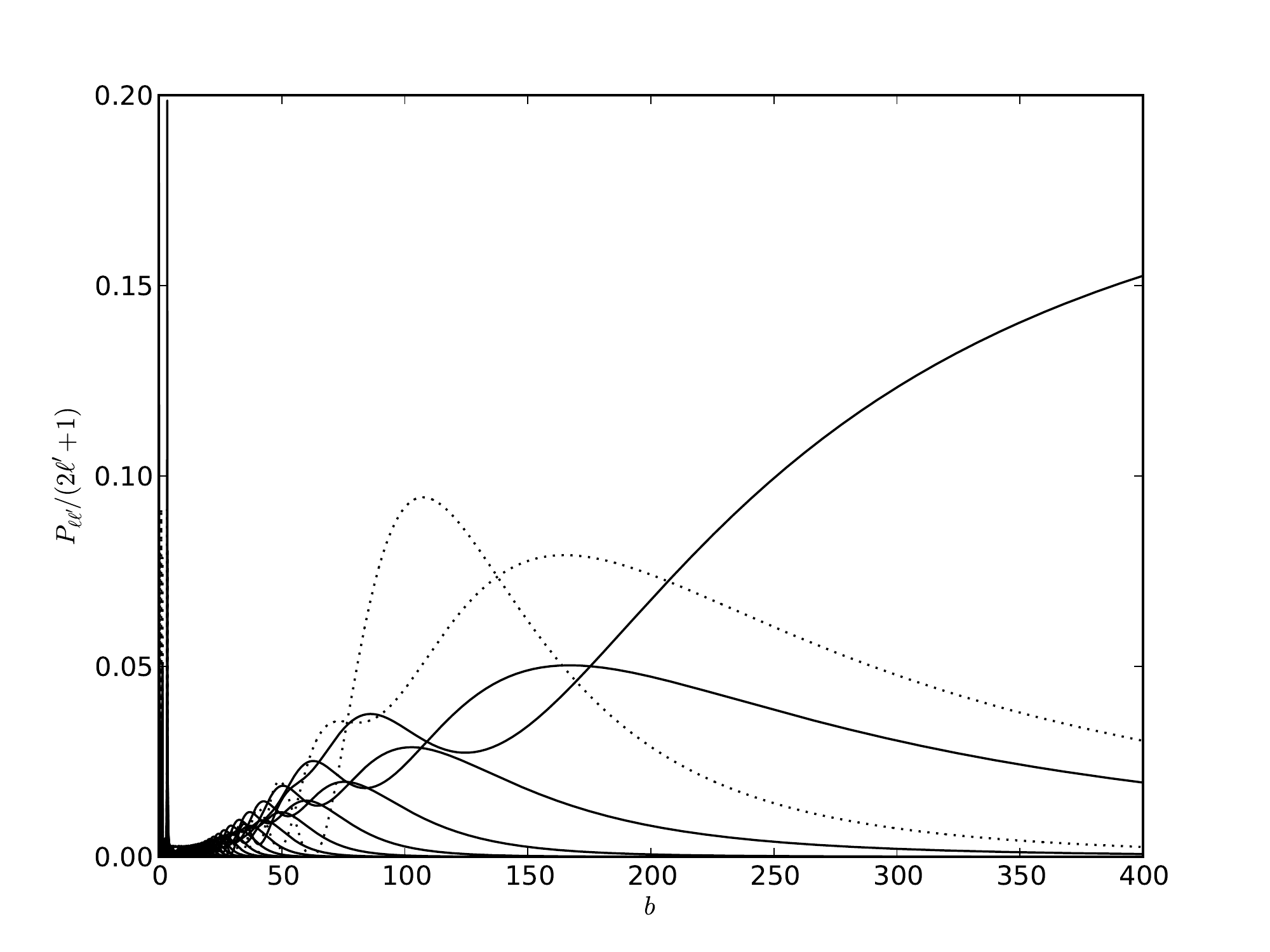} &
\includegraphics[width=0.4\textwidth]{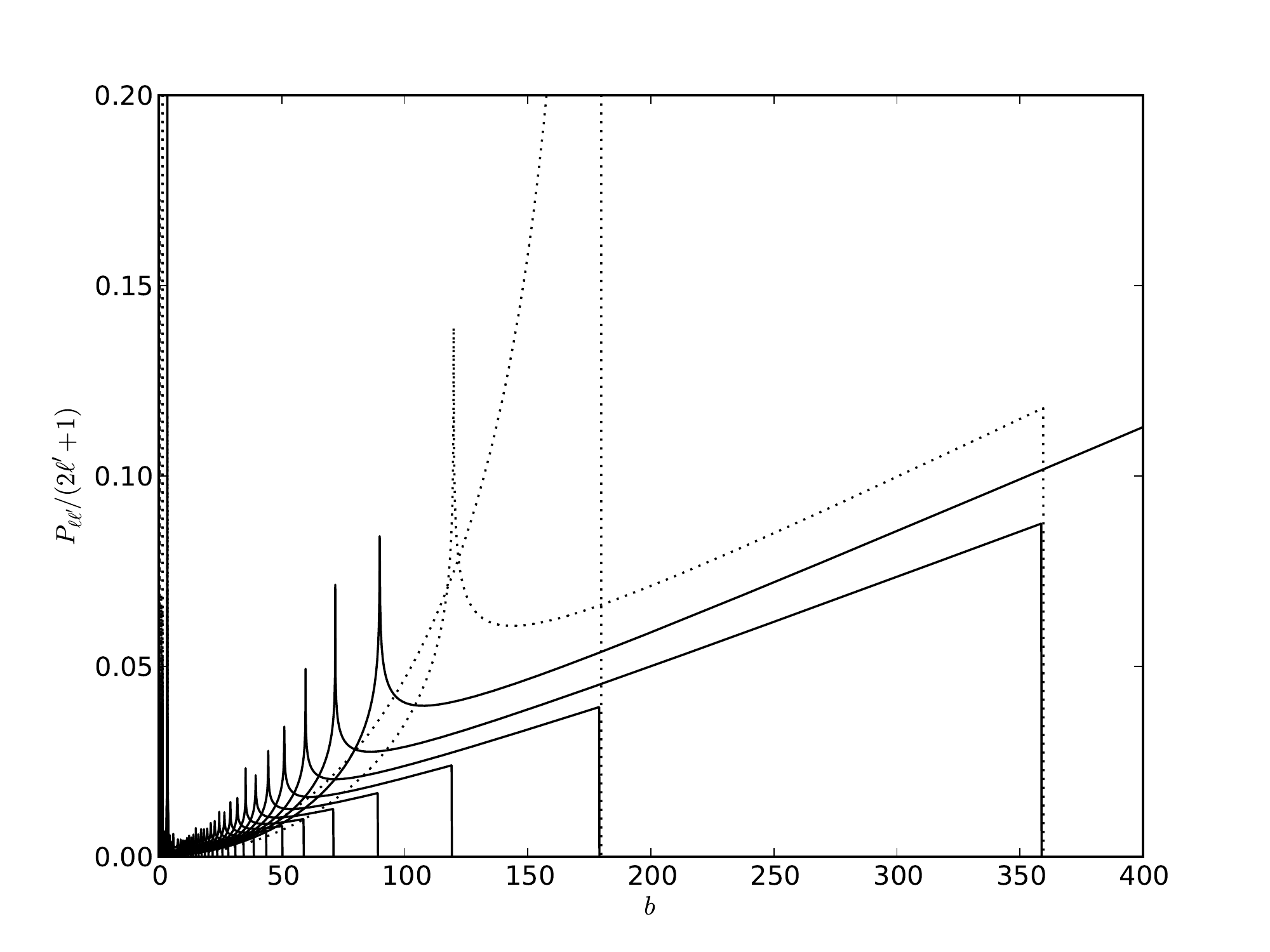} \\
\end{tabular}
\end{center}
\caption{Plots of $P_{n\ell\ell'}/(2\ell'+1)$ vs $b$ for $n=30$ and
  $\ell=0$ (first row), $\ell=1$ (second row) and $\ell=2$ (third
  row).  The Stark parameter is $\alpha = 6/b$.  Left column is using
  the QM formalism, right is using the SC formalism.  The highest
  curve at large $b$ has $\Delta\ell = 0$, larger $\Delta\ell$ curves
  appear in order as $b$ reduces.  Dotted curves are for $\ell' <
  \ell$.}
\label{f:sample1}
\end{figure*}
\begin{figure*}
\begin{center}
\begin{tabular}{cc}
\includegraphics[width=0.4\textwidth]{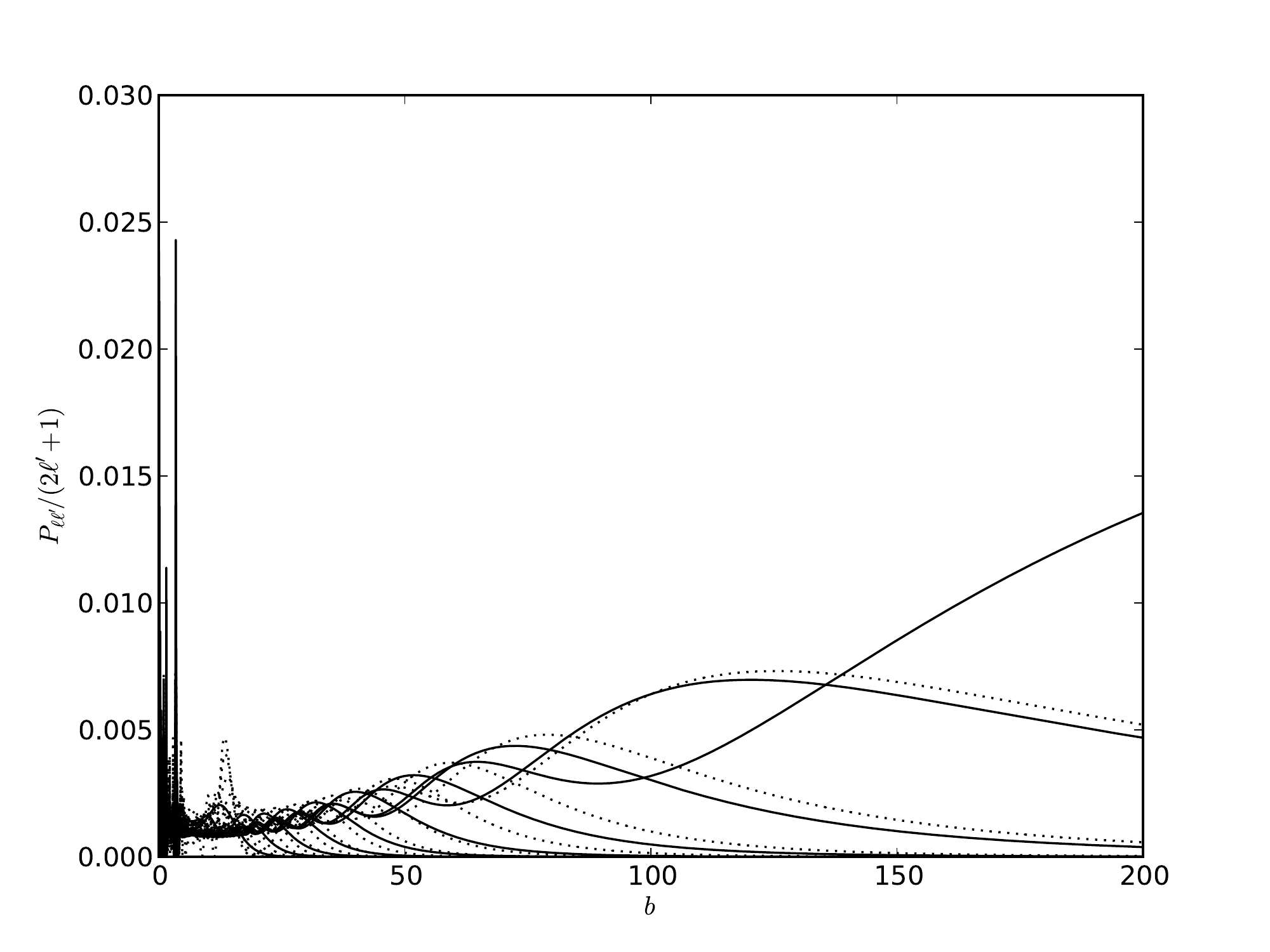} &
\includegraphics[width=0.4\textwidth]{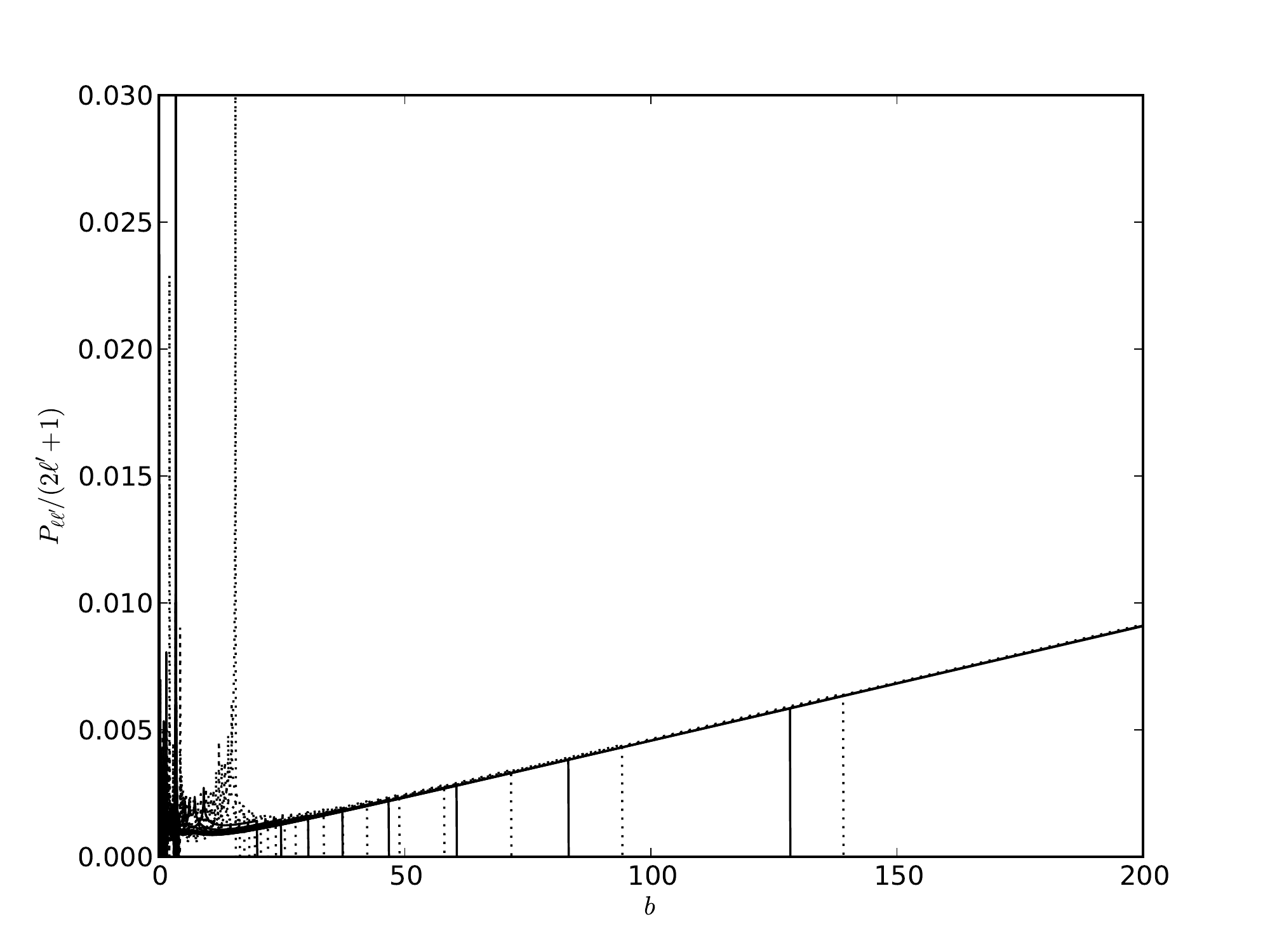} \\
\includegraphics[width=0.4\textwidth]{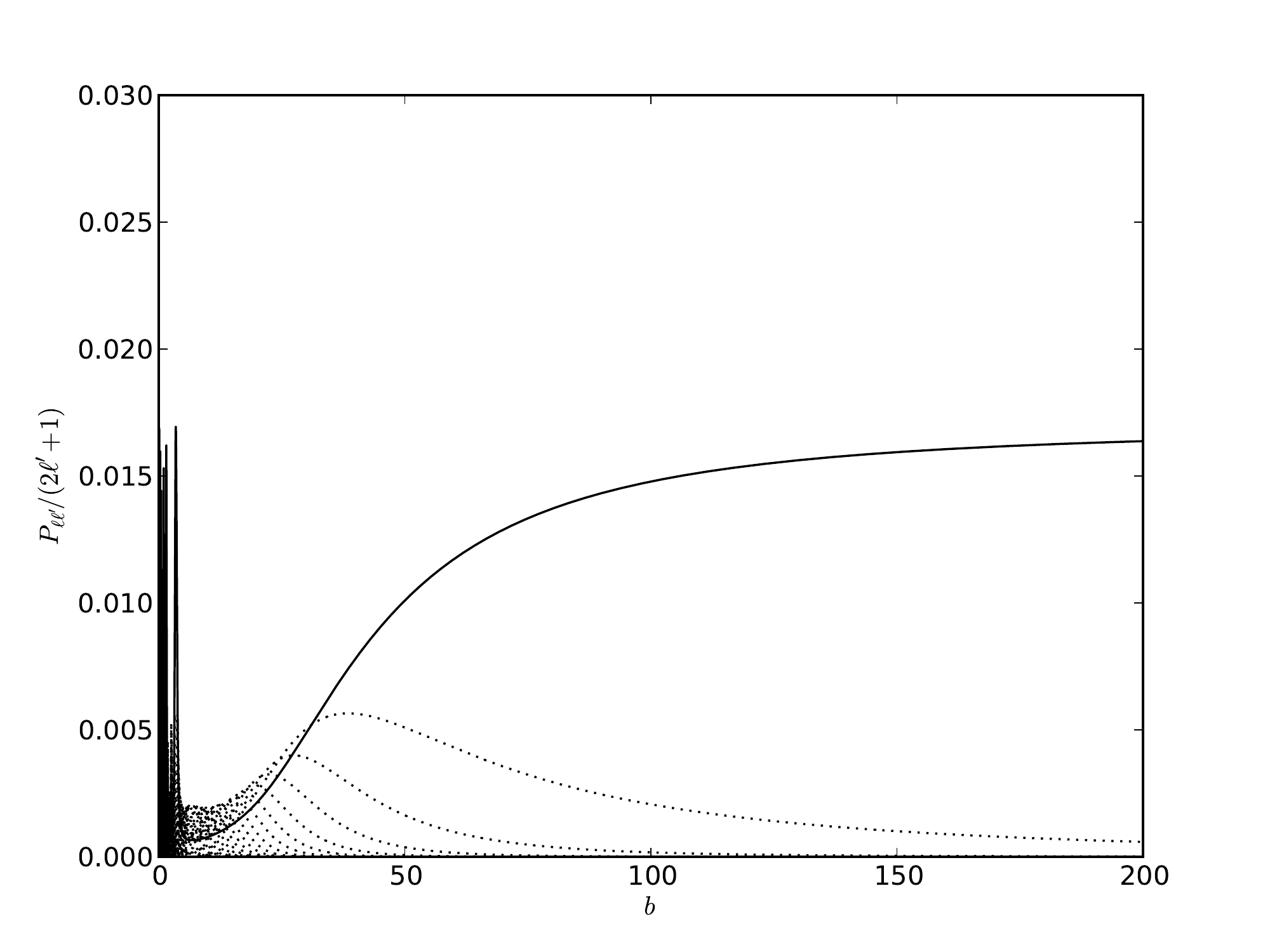} &
\includegraphics[width=0.4\textwidth]{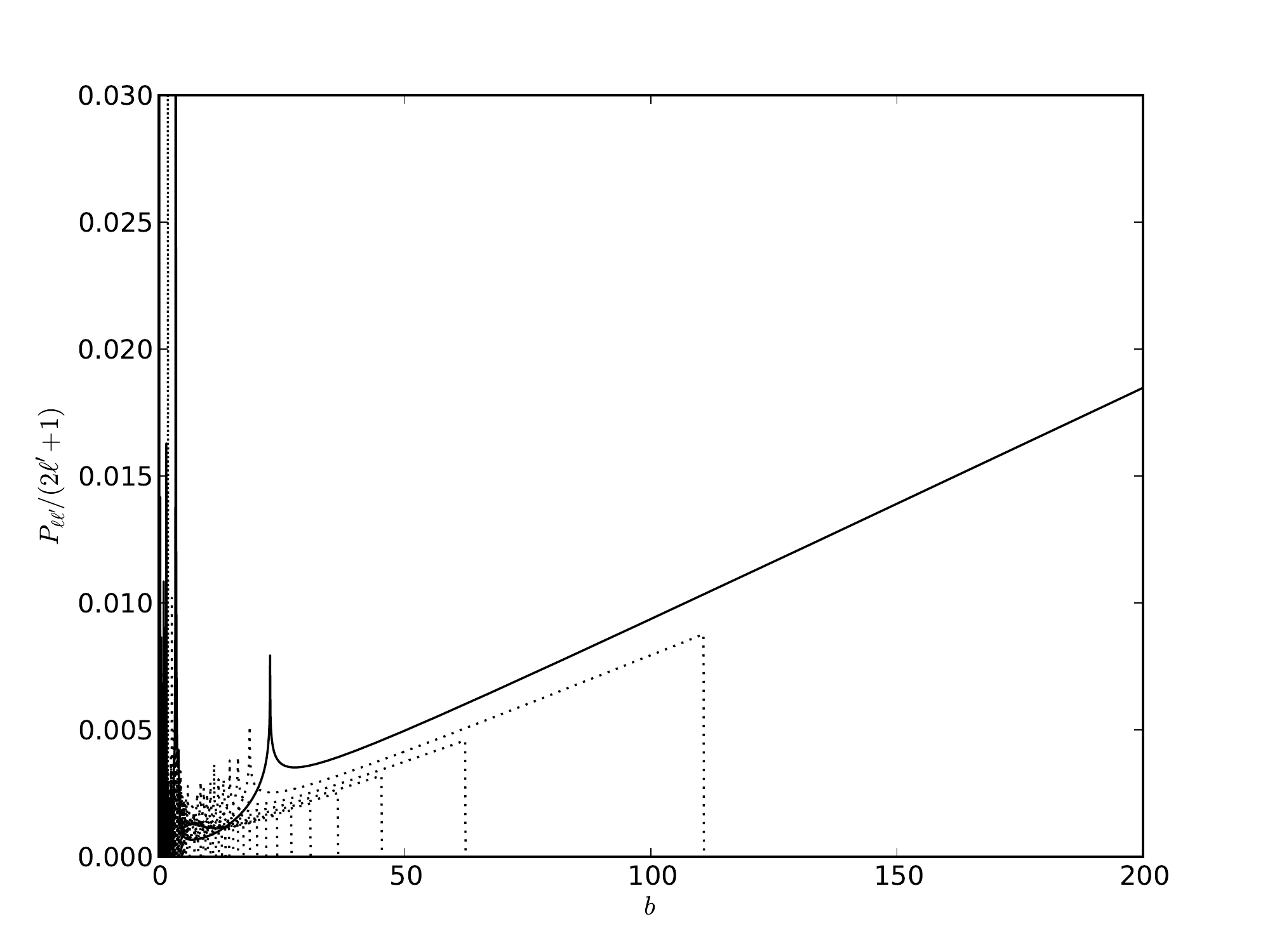}
\end{tabular}
\end{center}
\caption{Plots of $P_{n\ell\ell'}/(2\ell'+1)$ vs $b$ for $\ell=20$
  (first row) and $\ell=29$ (second row).  The $n$ and $\alpha$
  parameters are as in Figure~\ref{f:sample1}.  Left column is using
  the QM formalism, right is using the SC formalism.  Dotted curves
  are for $\ell' < \ell$.}
\label{f:sample2}
\end{figure*}

In Figures~\ref{f:sample1} and~\ref{f:sample2}, we compare the quantum
mechanical probability distributions with the classical transition
probabilities sampled at specific $\ell$, $\ell'$.  The QM dipole
transition rates, $\Delta\ell = \pm 1$ decay slowly as $b$ increases,
which is the origin of the divergence of the rate integral for these
transitions.  The classical transition probabilities show sharp edges
where the transitions first become allowed, for all $\Delta\ell$: the
transition rates for all $\Delta\ell$ are similar, as there is nothing
in the SC formulation which fundamentally distinguished a
$\vert\Delta\ell\vert=1$ transition from one with a larger change in
angular momentum.  There are also internal peaks for many cases,
corresponding to orbital resonances.  These are used by VOS12 to limit
the domain over which $\vert\Delta\ell\vert=1$ transitions are
allowed, avoiding the divergence in the integrated transition rate
found for the QM dipole transition rate.

It is clear that the classical transition probabilities cannot satisfy
unitarity, as where any transition ceases to be allowed, there is no
corresponding increase in the others.  Indeed, at sufficiently large
radii, the probability of no transition, $P_{n\ell\ell}$, increases
above unity, which is inconsistent with usual definition of
probability.  The quantum transition probabilities do satisfy
unitarity (note that the curves as plotted are divided by $2\ell'+1$
to make the ergodicity property at small $b$ more obvious, but this
means that this summation property for the probabilities is less
obvious as shown).

\begin{figure*}
\begin{center}
\begin{tabular}{cc}
\includegraphics[width=0.4\textwidth]{data0} &
\includegraphics[width=0.4\textwidth]{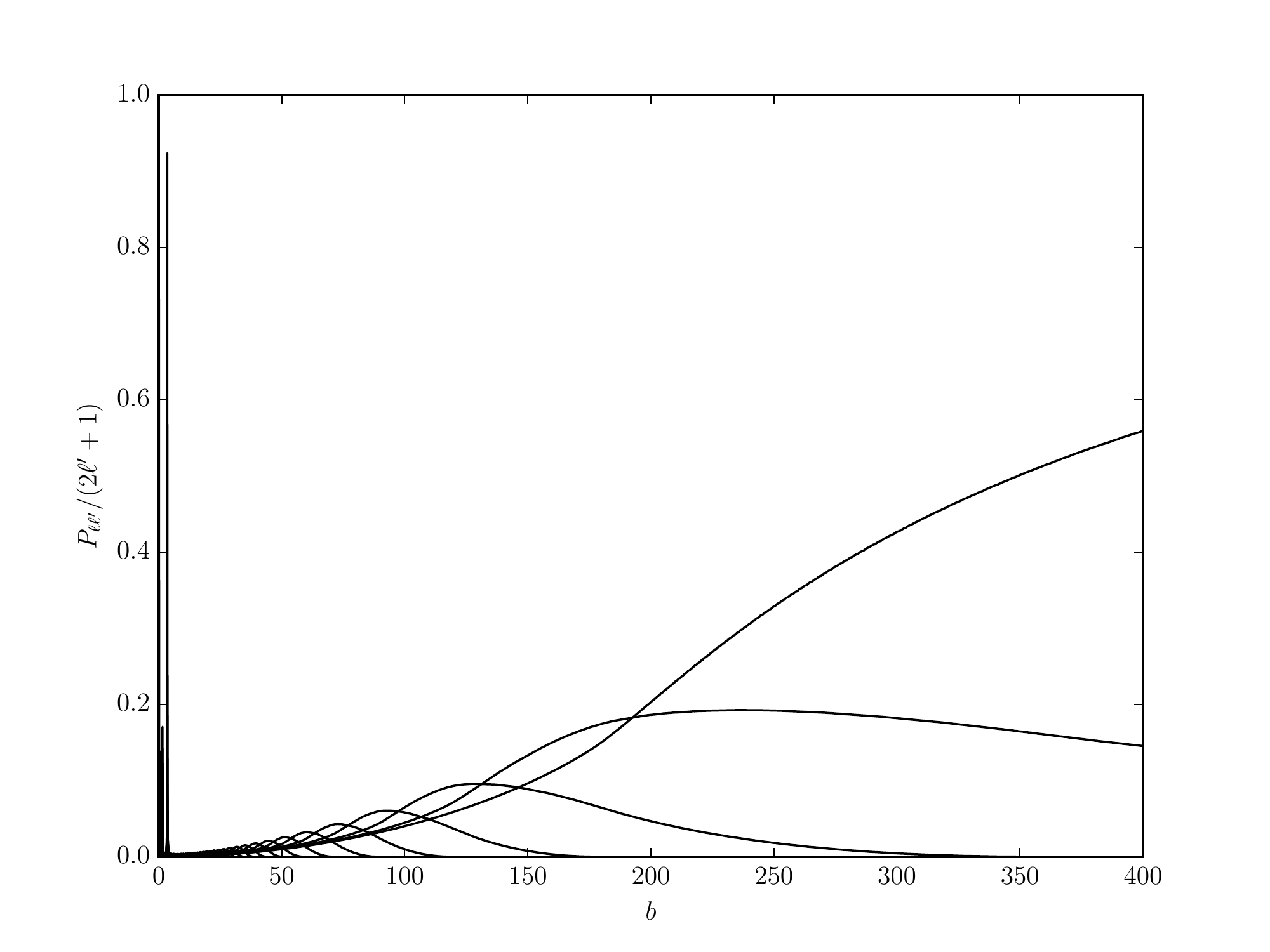} \\
\includegraphics[width=0.4\textwidth]{data1} &
\includegraphics[width=0.4\textwidth]{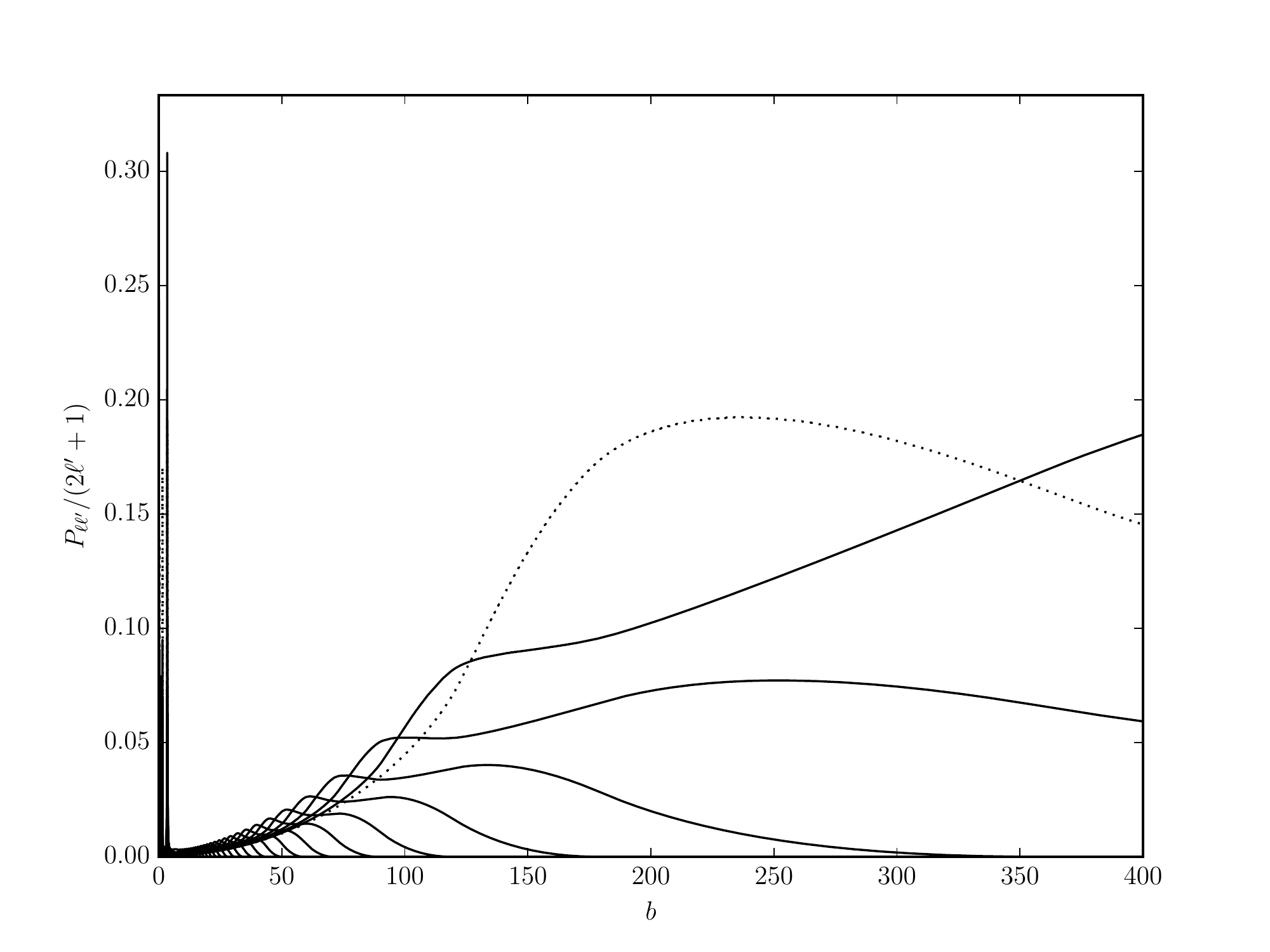} \\
\includegraphics[width=0.4\textwidth]{data2} &
\includegraphics[width=0.4\textwidth]{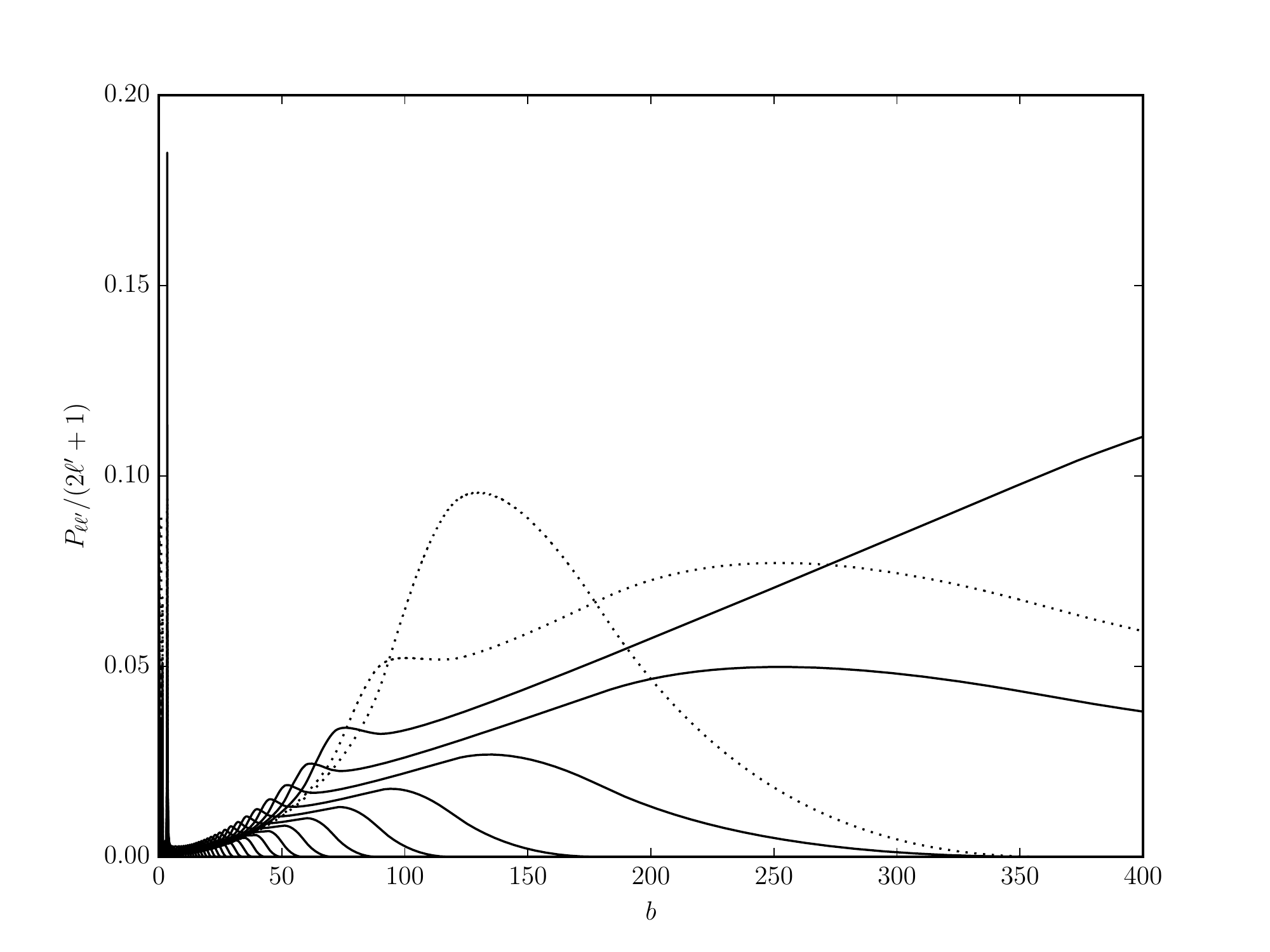} \\
\end{tabular}
\end{center}
\caption{Plots of $P_{n\ell\ell'}/(2\ell'+1)$ vs $b$ for $\ell=0$
  (first row), $\ell=1$ (second row) and $\ell=2$ (third row).  Left
  column is using the QM formalism, right is using the SC formalism,
  averaged over shells.  Dotted curves are for $\ell' < \ell$.}
\label{f:avg1}
\end{figure*}
\begin{figure*}
\begin{center}
\begin{tabular}{cc}
\includegraphics[width=0.4\textwidth]{data20} &
\includegraphics[width=0.4\textwidth]{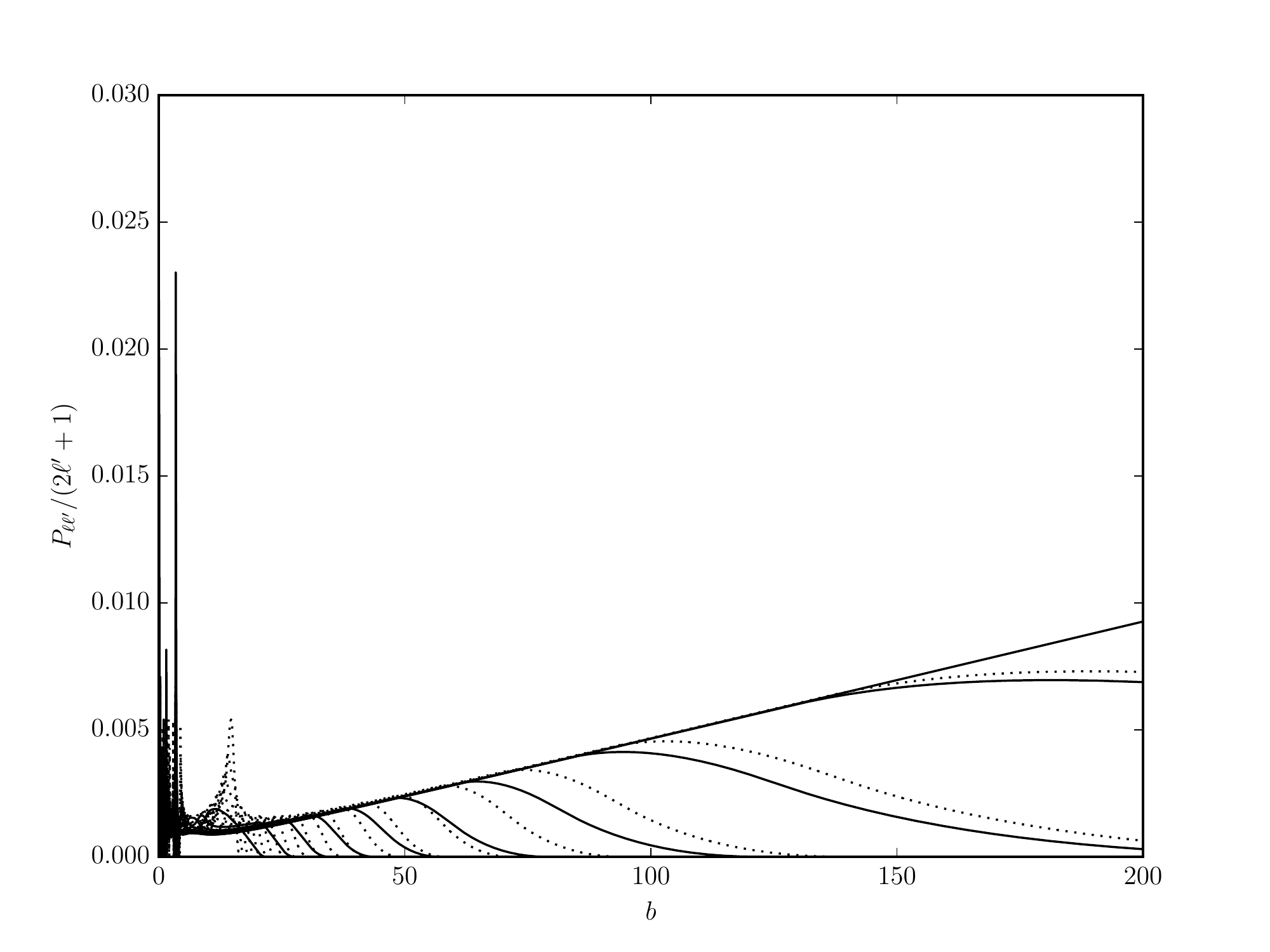} \\
\includegraphics[width=0.4\textwidth]{data29} &
\includegraphics[width=0.4\textwidth]{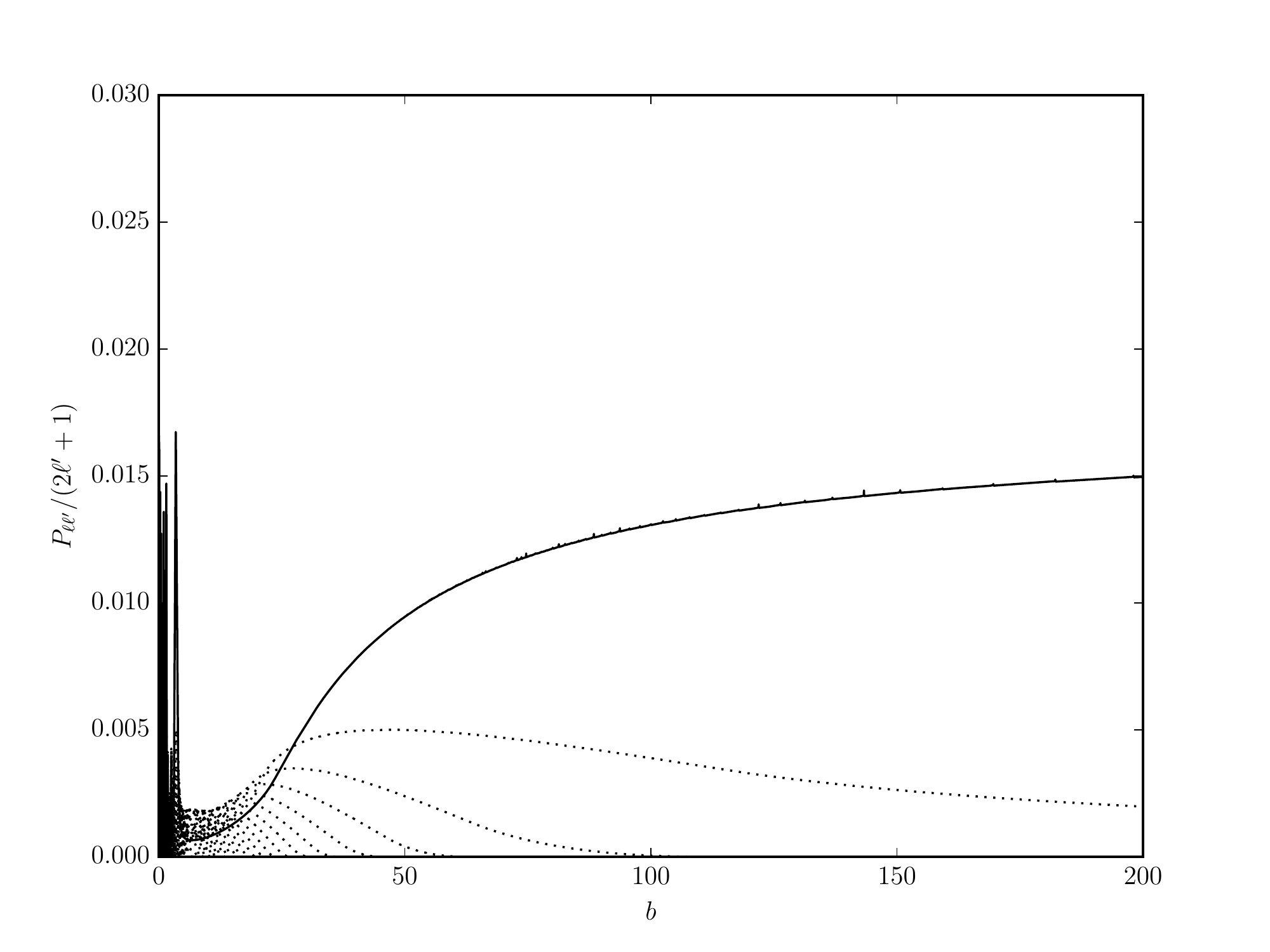}
\end{tabular}
\end{center}
\caption{Plots of $P_{n\ell\ell'}/(2\ell'+1)$ vs $b$ for $\ell=20$
  (first row) and $\ell=29$ (second row).  Left column is using the QM
  formalism, right is using the SC formalism, averaged over shells.
  Dotted curves are for $\ell' < \ell$.}
\label{f:avg2}
\end{figure*}

In Figures~\ref{f:avg1} and~\ref{f:avg2}, we compare the quantum
mechanical probability distributions with the classical probability
distributions averaged over angular momentum shells.  These plots are
significantly more alike than those for the comparison between the
quantum mechanical probability and the discretely-sampled classical
transition probability.  While there are no longer any sharp edges in
the shell-averaged classical probabilities, for $\vert\Delta\ell\vert
> 1$ they will be non-zero only within some range of $\alpha$ values.
(The integration over a quantized shell width means that there is now
a genuine distinction in qualitative behaviour between transitions
with different $\vert\Delta\ell\vert$.)  The shell-averaged classical
probabilities also satisfy unitarity and the quantum-weighted detailed
balance constraint.

Away from the region where the discretely-sampled transition
probability is zero, the binning has a relatively minor effect, simply
smoothing out the steepest peaks.

The general form of the transition probability distributions shown in
these figures is of interest.  Working from large $b$ inwards, it is
initially most likely that no change in $\ell$ will result.  At least
in the QM case, all other values of $\Delta\ell$ are possible, with
probabilities reducing as $\vert\Delta\ell\vert$ increases.  As $b$
becomes smaller, the probabilities of the higher
$\vert\Delta\ell\vert$ transitions increase, following a smooth power
law dependency, until the statisical weight of the output state
reaches a similar level to that of the $\Delta\ell=0$ transition.
Thereafter, the probabilities are subject to significant oscillations,
around an average level consistent with statistical balance, apart
from a strong spike at the very smallest values of $b$.  This suggests
that the combination of the asymptotic behaviour at large $b$, the
core ergodicity principle, and the fundamental requirements of
unitarity and detailed balance, should be sufficient to provide
thermodynamically-consistent estimates of the impact-parameter and
thermal-averaged rates which would be acceptably accurate for many
applications.

\section{Use of semi-classical probabilities}
\label{s:use}

The shell-sampled probabilities presented in the previous section are
determined using a computationally expensive double integral.  It may
be possible to perform one or both of these integrals analytically,
but numerical results were sufficient for the present analysis.

However, given that the use of a classical transition rate is already
a significant approximation, using a point sample of the transition
rate rather than an integral is likely to be an acceptable
approximation, at least away from the case of $\Delta\ell = \pm1$,
$b\to\infty$.  As these rates are being attributed to quantum
mechanical rather than classical states, it makes sense to ensure the
rates are chosen so as to satisfy quantum mechanical rather than
classical statistics.  The most consistent identification of quantum
mechanical states with a continuum band has $\ell\hbar$ as the angular
momentum at the innermost edge of the band.  Hence, if a single value
based on the classical transition probability is to be used for this
quantum number, it will be accurate to a higher order if the angular
momentum used in this expression is somewhat higher than $\ell\hbar$.
In particular, in order to be consistent with the quantum mechanical
detailed balance condition, the value $2\ell'+1$ should be used in the
numerator of the prefactor of equation (6) of VOS12.  This means that
the $\ell\to\ell'=0$ rates will not be strictly zero, as required by
the expressions given by VF01.  Given that VF01 provide expressions
for the transition rate {\em out}\/ of an $\ell=0$ state, a zero
inward rate is in clear violation of the detailed balance requirement.

The values used in the elliptic integral terms must be symmetric
functions of $\ell,\ell'$, as is true for the expressions given.
Ideally they should also be chosen to satisfy unitarity, but in
reality the correction to the overall transition rates as a result of
violating this constraint will be small compared to the other
approximations underlying this approach.  Using $\cos\eta_1 =
(\ell+{1\over2})/n$, etc., will be at least somewhat more accurate
than without the ${1\over2}$, and also means that the cases $\ell=0$,
$\ell'=0$ do not require a special treatment.

\section{Conclusions}
\label{s:concl}

We have shown that, by using an alternative form for the Monte Carlo
realization, the results for the SC and QM formalisms described by
VF01 and VOS12 can be brought closely into line.  As this is
consistent with what is expected as a result of the correspondence
principle, it provides further evidence that the results of the QM
formalism of VF01 and VOS12 should be preferred over their SC
formalism.

While finding agreement between the different forms of theory is
satisfying, this does not take into account the major reason given by
VF01 and VOS12 for preferring their semi-classical results,
specifically the need for an outer limit to be imposed on the
integration over impact parameters to prevent the total collision rate
diverging for $\Delta\ell = \pm 1$ when using the QM theory.  This
type of divergence is common in other areas of collision rate physics
(in particular, the two-body relaxation time,
\cite{1987degc.book.....S,1989islp.book.....M}), so is not unexpected,
and the empirical limits of PS64 are similar to those applied to the
Coulomb logarithm in these contexts.

However, as $b$ increases, so does the time over which the collision
takes place: the treatment of collisions as independent events must
therefore eventually become inaccurate.  Following the reasoning
underlying the diffusion-based approach to modelling transition rates
between degenerate levels discussed by \cite{1979JPhB...12.2051P}
(which requires no empirical cut-offs), we note that during the
extended period taken to complete the most distant encounters, there
will potentially be time for many collisions at smaller impact
parameters.  While, at first order, the effects of these collisions
will superpose linearly, for collisions at sufficiently large $b$, a
limit will be reached where the smaller impact parameter collisions
together are sufficient to randomize the angular momentum of the
target orbital.  Once $b$ increases above the level where this occurs,
$b_{\rm eq}$, the effect of collisions at larger impact parameter will
be felt, in effect, as a superposition of $N_{\rm{}coll}\sim
\tau_{\rm{}coll}/\tau_{\rm{}eq}$ partial interactions adding in
quadrature, rather than linearly (where $\tau_{\rm{}coll}$ is the
collision time at the large $b$ of interest, and $\tau_{\rm{}eq}$ is
the collision time at the smallest radius leading to effective
randomization).  A reduction in contribution to the transition
probability by $\sim N_{\rm{}coll}^{-1/2} \propto b^{-1/2}$ at the
largest $b$ will be sufficient to prevent the weak logarithmic
divergence in the overall rate.  This is a somewhat academic argument,
as radiative lifetimes and plasma particle correlations of the type
discussed by PS64 will often result in more stringent limits to the
range of $b$ over which collisions are effective.  Nevertheless, given
that the agreement we now find between the SC and QM gives greater
confidence in SC results for the Rydberg scattering problem, it may be
possible to investigate the corrections required for these multiple
interactions in a believable manner using explicit classical
trajectory calculations.

\section*{Acknowledgements}

We thank D. Vrinceanu and H. Sadeghpour for helpful responses to
several queries about their published work.  Parts of this work have
been supported by the NSF (1108928, 1109061, and 1412155), NASA
(10-ATP10-0053, 10-ADAP10-0073, NNX12AH73G, and ATP13-0153), and STScI
(HST-AR-13245, GO-12560, HST-GO-12309, GO-13310.002-A, and
HST-AR-13914). MC has been supported by STScI (HST-AR-14286.001-A).
PvH was funded by the Belgian Science Policy Office under contract
no.\ BR/154/PI/MOLPLAN.

\bibliography{vosc}		

\begin{thebibliography}{10}

\bibitem{2010MNRAS.407..599C}
J.~{Chluba}, G.~M. {Vasil}, and L.~J. {Dursi}.
\newblock {Recombinations to the Rydberg states of hydrogen and their effect
  during the cosmological recombination epoch}.
\newblock {\em \mnras}, 407:599--612, September 2010.

\bibitem{Vrinceanu2001b}
D.~{Vrinceanu} and M.~R. {Flannery}.
\newblock {LETTER TO THE EDITOR: Analytical quantal collisional Stark mixing
  probabilities}.
\newblock {\em Journal of Physics B Atomic Molecular Physics}, 34:L1--L8,
  January 2001.

\bibitem{VOS2012}
D.~{Vrinceanu}, R.~{Onofrio}, and H.~R. {Sadeghpour}.
\newblock {Angular Momentum Changing Transitions in Proton-Rydberg Hydrogen
  Atom Collisions}.
\newblock {\em \apj}, 747:56, March 2012.

\bibitem{2016MNRAS.459.3498G}
F.~{Guzm{\'a}n}, N.~R. {Badnell}, R.~J.~R. {Williams}, P.~A.~M. {van Hoof},
  M.~{Chatzikos}, and G.~J. {Ferland}.
\newblock {H, He-like recombination spectra - I. l-changing collisions for
  hydrogen}.
\newblock {\em \mnras}, 459:3498--3504, July 2016.

\bibitem{2017MNRAS.464..312G}
F.~{Guzm{\'a}n}, N.~R. {Badnell}, R.~J.~R. {Williams}, P.~A.~M. {van Hoof},
  M.~{Chatzikos}, and G.~J. {Ferland}.
\newblock {H-, He-like recombination spectra - II.l-changing collisions for He
  Rydberg states}.
\newblock {\em \mnras}, 464:312--320, January 2017.

\bibitem{1984JPhB...17.3923B}
R.~L. {Becker} and A.~D. {MacKellar}.
\newblock {Theoretical initial l dependence of ion-Rydberg-atom collision cross
  sections}.
\newblock {\em Journal of Physics B Atomic Molecular Physics}, 17:3923--3942,
  October 1984.

\bibitem{leveque2002finite}
Randall~J LeVeque.
\newblock {\em Finite volume methods for hyperbolic problems}, volume~31.
\newblock Cambridge {U}niversity {P}ress, 2002.

\bibitem{landau2014guide}
David~P Landau and Kurt Binder.
\newblock {\em A guide to {M}onte {C}arlo simulations in statistical physics}.
\newblock {C}ambridge {U}niversity {P}ress, 2014.

\bibitem{1967MNRAS.136..101L}
D.~{Lynden-Bell}.
\newblock {Statistical mechanics of violent relaxation in stellar systems}.
\newblock {\em \mnras}, 136:101, 1967.

\bibitem{1964MNRAS.127..165P}
R.~M. {Pengelly} and M.~J. {Seaton}.
\newblock {Recombination spectra, II}.
\newblock {\em \mnras}, 127:165, 1964.

\bibitem{1987degc.book.....S}
L.~{Spitzer}.
\newblock {\em {Dynamical evolution of globular clusters}}.
\newblock Princeton, NJ, Princeton University Press, 1987.

\bibitem{1989islp.book.....M}
D.~B. {Melrose}.
\newblock {\em {Instabilities in Space and Laboratory Plasmas}}.
\newblock Cambridge, UK: Cambridge University Press, March 1989.

\bibitem{1979JPhB...12.2051P}
I.~C. {Percival} and D.~{Richards}.
\newblock {Classical theory of transitions between degenerate states of excited
  hydrogen atoms in plasma}.
\newblock {\em Journal of Physics B Atomic Molecular Physics}, 12:2051--2065,
  June 1979.

\end{thebibliography}

\label{lastpage}
\clearpage
\end{document}